\documentclass[11pt]{article}
\usepackage[margin=1in]{geometry}
\usepackage{graphicx} 
\usepackage{titlesec}
\usepackage{multirow}%
\usepackage{url}
\usepackage{mathtools}
\usepackage{amsmath}
\usepackage{array}
\usepackage{tabularx}
\usepackage{listings}[R]
\usepackage[raggedrightboxes]{ragged2e}
\usepackage{xcolor}
\usepackage{amsthm}
\usepackage{textcomp}%
\usepackage{manyfoot}%
\usepackage{enumitem}
\usepackage{bm}
\usepackage{array}
\usepackage[font=small]{caption}

\usepackage{booktabs}%
\usepackage{setspace}
\usepackage[colorlinks=true,
            linkcolor=blue,
            urlcolor=blue,
            citecolor=blue]{hyperref}

\usepackage{natbib}

\titleformat{\section}
  {\normalfont\itshape\large}
  {\thesection}
  {1em}
  {}

\titleformat{\subsection}
  {\normalfont\itshape\normalsize}
  {\thesubsection}
  {1em}
  {}

\titleformat{\subsubsection}
  {\normalfont\itshape\normalsize}
  {\thesubsubsection}
  {1em}
  {}

\titleformat{\paragraph}[runin]
  {\normalfont\itshape\normalsize}
  {\theparagraph}
  {1em}
  {}
\titlespacing*{\paragraph}{0pt}{1ex}{1em}

\newcommand{\lm}{\texttt{robin\_lm }}
\newcommand{\glm}{\texttt{robin\_glm }}

\title{The RobinCar Family: R Tools for Robust Covariate Adjustment in Randomized Clinical Trials}
\author{
Marlena Bannick$^{1\dagger}$,
Yuanyuan Bian$^{2\dagger}$,
Gregory Chen$^{3\dagger}$,
Liming Li$^{4\dagger}$,\\
Yuhan Qian$^{1\dagger}$,
Daniel Sabanés Bové$^{5\dagger}$,
Dong Xi$^{6\dagger}$,
Ting Ye$^{1\dagger}$,
Yanyao Yi$^{2\dagger}$ \vspace{3mm}\\
on behalf of the Software Subteam\\
ASA-BIOP Covariate Adjustment Scientific Working Group \vspace{3mm} \\
$^\dagger$All authors contributed equally to this work.
\vspace{3mm} \\
$^{1}$Department of Biostatistics, University of Washington, Seattle, WA, USA\\
$^{2}$Global Statistical Sciences, Eli Lilly and Company, Indianapolis, Indiana, USA\\
$^{3}$BARDS, MSD, Switzerland\\
$^{4}$AstraZeneca\\
$^{5}$RCONIS\\
$^{6}$Gilead Sciences\\
}

\begin{document}

\maketitle

\abstract{\textbf{Purpose:} Covariate adjustment is a powerful statistical technique that can increase efficiency in clinical trials. Recent guidance from the U.S. FDA provided recommendations and best practices for using covariate adjustment. However, there has existed a gap between the extensive statistical literature on covariate adjustment and software that is easy to use and abides by these best practices.

\textbf{Methods:} We have developed the RobinCar Family, which is comprised of RobinCar and RobinCar2. These two R packages enable covariate-adjusted analyses for continuous, discrete, and time-to-event outcomes that follow best practices. For continuous and discrete outcomes, the functions in the RobinCar Family facilitate traditional forms of covariate adjustment such as ANCOVA as well as more recent approaches like ANHECOVA, G-computation with generalized linear models and machine learning models, and adjustment for a super-covariate (as in PROCOVA\texttrademark). Functions for time-to-event outcomes implement the covariate-adjusted log-rank test, the stratified covariate-adjusted log-rank test, and the marginal covariate-adjusted hazard ratio. The RobinCar Family is supported by the ASA Biopharmaceutical Section Covariate Adjustment Scientific Working Group.
 
\textbf{Results:} We provide an accessible overview of the covariate-adjusted statistical methods, and describe how they are implemented in RobinCar and RobinCar2. We highlight important usage notes for clinical trial practitioners.
 
\textbf{Conclusion:} We apply RobinCar and RobinCar2 functions by analyzing data from the AIDS Clinical Trials Group Study 175, demonstrating that they are straightforward and user-friendly.}

\vspace{3mm}


\section{Background}\label{sec1}

Randomized clinical trials are the foundation of successful drug and biological product development, but they are also resource-intensive.
Therefore, efficiency is of paramount concern. Covariate adjustment is a statistical technique that utilizes prognostic pre-randomization variables to improve efficiency for demonstrating and quantifying treatment effects. When done correctly, covariate adjustment can decrease the variance of estimated intervention effects, leading to more precise conclusions or reducing the number of patients needed to maintain power as compared to a so-called ``un-adjusted'' analysis that does not incorporate covariates.

There is extensive statistical literature on robust covariate adjustment using linear working models (e.g., \cite{yangEfficiencyStudyEstimators2001, tsiatisCovariateAdjustmentTwosample2008, freedmanRegressionAdjustmentsExperimental2008, linAgnosticNotesRegression2013, jiangRobustAlternativesANCOVA2019, bugniInferenceCovariateAdaptiveRandomization2018, bugniInferenceCovariateadaptiveRandomization2019, liRerandomizationRegressionAdjustment2020, yeInferenceAverageTreatment2022, yeBetterPracticeCovariate2023}), non-linear working models (e.g., \cite{freedmanRandomizationDoesNot2008, mooreCovariateAdjustmentRandomized2009, wangModelRobustInferenceClinical2023, guoGeneralizedOaxacaBlinderEstimator2023, cohenNoharmCalibrationGeneralized2024, liu_covariate_2024}), and for time-to-event outcomes (e.g., \cite{luImprovingEfficiencyLogrank2008, mooreIncreasingPowerRandomized2009, yeCovariateAdjustedLogRankTest2023, yeRobustTestsTreatment2020}). Despite its potential benefits, covariate adjustment is still underutilized in clinical trials. One major obstacle is that new methods and software (when available) are scattered throughout the literature, and they are often limited to special situations making it difficult to validate each separate software and inconvenient to use in practice (see Table \ref{tab:software}). This has resulted in current trial analyses either unnecessarily relying on modeling assumptions or simply forgoing covariate adjustment (and thus losing potential efficiency gain). This is one example of missing commercialization of a statistical invention \citep{rufibachImplementationStatisticalInnovation2025}.

\begin{table}
\small
\centering
\caption{Existing \texttt{R} CRAN packages for covariate adjustment in randomized clinical trials.}\label{tab:software}
\begin{tabular}{p{4cm} | p{3cm} p{5cm}}
Name & Citation & Description \\
& & \\
\hline
\hline
& & \\
\texttt{RobinCar}: Robust Inference for Covariate Adjustment in Randomized Clinical Trials & \cite{bannickRobinCarRobustInference2024} & comprehensive package on covariate adjustment in randomized clinical trials for continuous, binary, and time-to-event outcomes, including inference under covariate-adaptive randomization \\
& & \\
\hline
& & \\
\texttt{RobinCar2}: ROBust INference for Covariate Adjustment in Randomized Clinical Trials & \cite{liRobinCar2ROBustINference2025} & lite and well-validated version of \texttt{RobinCar} for regulatory purposes \\
& & \\
\hline
& & \\
\texttt{beeca}: Binary Endpoint Estimation with Covariate Adjustment
& \cite{przybylskiBeecaBinaryEndpoint2024} & estimation of marginal treatment effects using G-computation based on logistic regression model in simple randomization, provides variance estimators from \cite{geCovariateAdjustedDifferenceProportions2011} and \cite{yeRobustVarianceEstimation2023} \\
& & \\
\hline
& & \\
\texttt{mmrm}: Mixed Models for Repeated Measures & \cite{sabanesboveMmrmMixedModels2024} & for analyzing longitudinal continuous outcomes in RCTs using a linear model \\
& & \\
\hline
& & \\
\texttt{AIPW}: Augmented Inverse Probability Weighting & \cite{zhongAIPWAugmentedInverse2025} & augmented inverse probability weighted-estimators -- package is for both experimental and observational studies \\
& & \\
\hline
& & \\
\texttt{carat}: An R Package for Covariate-Adaptive Randomization in Clinical Trials & \cite{tuCaratCovariateAdaptiveRandomization2023} & provides functions to perform hypothesis tests on data from trials that use covariate-adaptive randomization, including randomization-based testing \\
& & \\
\hline
\end{tabular}
\end{table}

To address the gap between advancement in statistical methods and clinical trial practice, the Food and Drug Administration (FDA) released a timely guidance for industry on “Adjustment for Covariates in Randomized Clinical Trials for Drugs and Biological Products” \citep{u.s.foodanddrugadministrationAdjustingCovariatesRandomized2024} to provide clarifications and recommendations on the use of covariate adjustment. In response, there has been increasing interest among clinical trial practitioners in using covariate adjustment. Consequently, there is a pressing need for easy-to-use, comprehensive, and reliable software for covariate adjustment in the statistical analysis of randomized clinical trials. In addition to obtaining covariate-adjusted estimators, the software should also provide variance estimators that are robust against model misspecification and heteroscedasticity, and that account for commonly-used covariate-adaptive randomization schemes such as stratified permuted block randomization \citep{zelenRandomizationStratificationPatients1974} and Pocock-Simon minimization schemes \citep{pocockSequentialTreatmentAssignment1975}. 

To meet this need, we have developed a family of R packages for covariate adjusted analyses called the RobinCar Family. The RobinCar Family allows clinical trial practitioners to easily apply covariate adjustment to analyze continuous, discrete, and time-to-event outcomes under simple, biased coin, stratified permuted block, and Pocock-Simon's minimization randomization. Our work provides a much-needed, unified platform to implement covariate adjustment in accordance with FDA's guidance and enables more efficient analysis of randomized clinical trial data without compromising robustness.

\section{Implementation}

\texttt{RobinCar} was initially developed to provide R users with access to new and promising covariate adjustment methodologies. It is a one-stop package that incorporates covariate adjustment methodologies for continuous, discrete, and time-to-event outcomes. As interest in the statistical methods available in \texttt{RobinCar} grew, industry sponsors sought to utilize \texttt{RobinCar} to analyze their clinical trial data, and ultimately to provide evidence for their applications to regulators. At the same time, the American Statistical Association's (ASA) Biopharmaceutical Section was forming a working group on covariate adjustment, including a sub-team specifically focused on software development. This provided a unique opportunity to develop a new and streamlined package, \texttt{RobinCar2}, with a broader array of stakeholders including academic researchers, industry professionals, and regulators. The goal of establishing \texttt{RobinCar2} is to provide clinical trial practitioners with a lite version of \texttt{RobinCar} that has only the most essential and well-validated statistical methods, following good practices for engineering high quality statistical software packages \citep{openstatsguideOpenstatsguideMinimumViable2024}.

\subsection{RobinCar Family Principles}

The covariate adjustment methods that are available in the RobinCar Family are driven by the following principles:

\begin{itemize}[label=]
	\item \textit{\textbf{Estimand-Focused}: The estimand (target of estimation) should be determined by the trial objective, and the choice of estimator should be driven by the estimand. Covariate-adjusted estimators and hypothesis tests should be appropriate for the estimand or null hypothesis that were determined by the trial objective.}
	\item \textit{\textbf{Assumption-Lean}: Covariate-adjusted methods should be valid ``under the same minimal statistical assumptions that would be needed for unadjusted estimation'' \citep{u.s.foodanddrugadministrationAdjustingCovariatesRandomized2024} For example, covariate adjustment methods that rely on a working model should not require correct model specification for point estimates to be consistent, or for confidence intervals to have nominal coverage.}
	\item \textit{\textbf{Fit-for-Purpose Variance Estimator}: Variance estimators for covariate adjustment should be robust to model misspecification and heteroscedasticity, and also be appropriate for the randomization scheme that was used in the trial (e.g.,  stratified permuted block randomization).}
\end{itemize}
The above principles are driven both by the FDA Guidance on Covariate Adjustment and the ICH E9(R1) Addendum on Estimands and Sensitivity Analysis in Clinical Trials. The essence of the three principles is that the choice to use a covariate-adjusted estimator should not change the question of interest or rely on strong assumptions, and should be paired with fully robust variance estimation methods.

One additional desirable property is that covariate adjustment should improve efficiency. That is, the variance of the covariate-adjusted estimator should be less than the variance of an un-adjusted estimator that targets the same estimand. In some cases, efficiency gain is guaranteed if prognostic baseline variables are used. For example, ANHECOVA, or ANCOVA2, which uses a linear model with treatment-by-covariate interactions, is guaranteed to be asymptotically more efficient than ANOVA \citep{yeBetterPracticeCovariate2023}. However, such gains may not materialize in finite samples. In other cases, efficiency gain is not straightforward and it is related to how well a ``working model'' encodes the true relationship between the prognostic variables and the outcome of interest, which requires domain-specific knowledge or practical strategies that will be detailed in the coming sections. As such, the RobinCar Family allows flexible specification of covariate adjustment methods, not limited to only those with theoretically guaranteed efficiency gains.

\subsection{The RobinCar Family Lifecycle}

\begin{figure}[htbp!]
	\caption{The RobinCar Family Lifecycle}\label{fig:robin-lifecycle}
	\includegraphics[width=\textwidth]{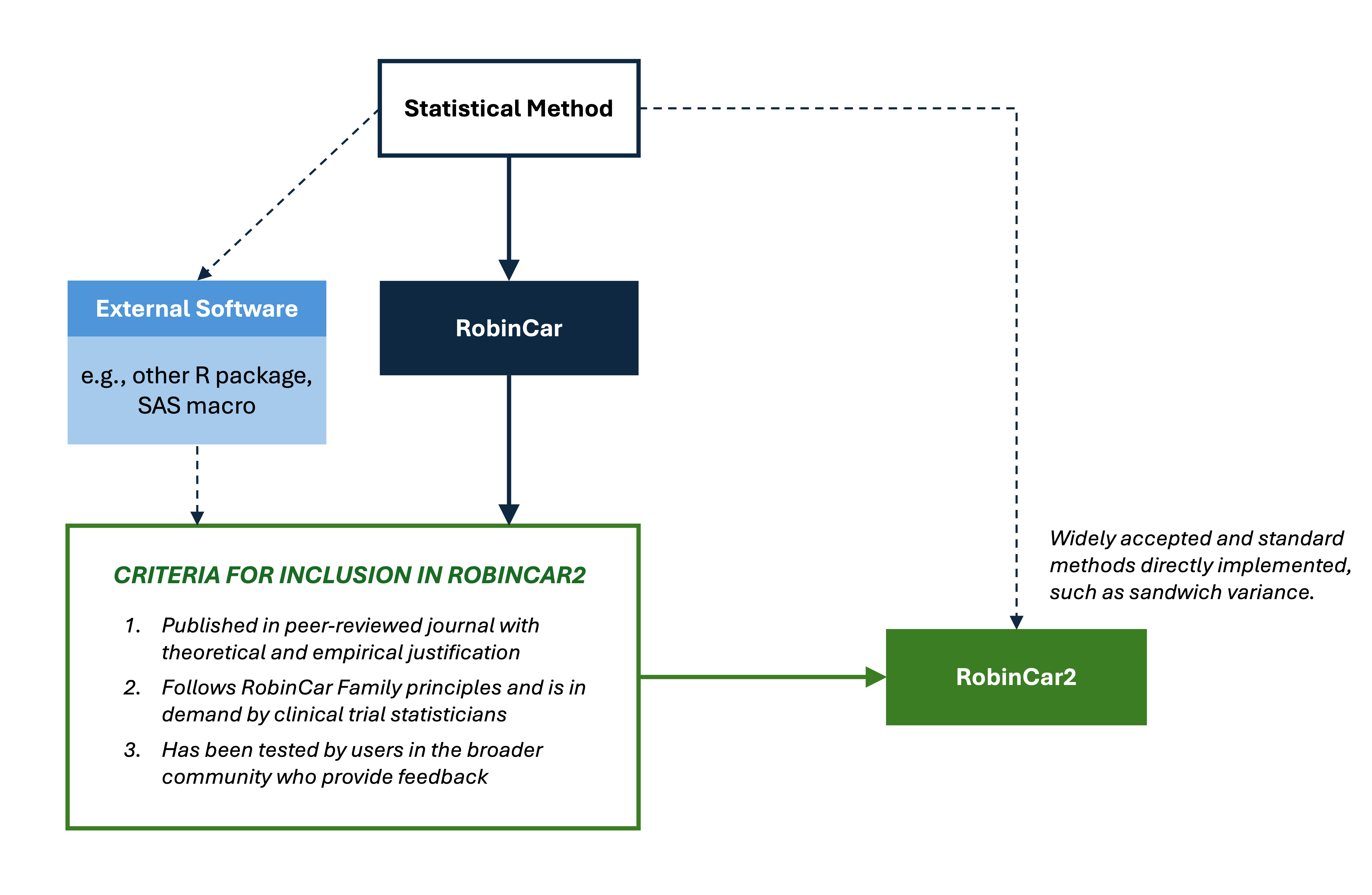}
\end{figure}

In Figure \ref{fig:robin-lifecycle}, we provide an overview of the lifecycle of a covariate adjustment method within the RobinCar Family.

When a new covariate adjustment methodology for RCTs is presented in the statistical literature and it abides by the RobinCar principles, it can be incorporated into the \texttt{RobinCar} package. Academics and industry researchers may provide feedback to the authors, which initiates an iterative process of improving the user-facing functions, as well as the underlying statistical methodology.

If the statistical method meets key criteria, it can further be incorporated into \texttt{RobinCar2}. The first is that the method must be published (or accepted for publication) in a peer-reviewed journal, which is accompanied by both theoretical and empirical justifications. In particular, the publication must demonstrate satisfactory performance of the method in realistic simulation studies of randomized clinical trials. The second criterion is that the method is in line with RobinCar Family principles, and is in demand by clinical trials statisticians and practitioners. The final criterion is that the method has been tested by users in the broader community who can provide feedback on the usability of the function(s), and can identify issues with the implementation of the method when it is applied to real data analyses.

After a statistical method has been implemented into \texttt{RobinCar2}, users of \texttt{RobinCar} that use a function that exists in both \texttt{RobinCar} and \texttt{RobinCar2} are met with a message to consider using \texttt{RobinCar2} instead. Although the internal methodology is identical, \texttt{RobinCar2} is supported and endorsed by the Software Sub-team of the American Statistical Association Covariate Adjustment Scientific Working Group. Therefore, for long-term stability and maintenance purposes, \texttt{RobinCar2} is the preferred choice, and is recommended for regulatory purposes.

There are two other ways that statistical methods may be incorporated into \texttt{RobinCar2} without having to first be in \texttt{RobinCar}. The first is if there are statistical methods incorporated into other less-standardized R packages, or publicly-available SAS macros. These may also meet the criteria for inclusion in Figure \ref{fig:robin-lifecycle}, and can be directly incorporated into \texttt{RobinCar2}. Finally, if there is a widely used and standard statistical method, it may be included in \texttt{RobinCar2} (e.g., sandwich standard errors). However, because methods need to be fully robust, these methods may only be available in specific circumstances \citep{yeBetterPracticeCovariate2023}. For example, in the context of the FDA guidance document for linear contrasts from a linear model, sandwich standard errors are only valid when treatment-by-covariate interactions are not used (i.e., ANCOVA).

\section{Results}

Table \ref{tab:robincar-methods} provides an overview of the methodologies that are available in \texttt{RobinCar} and \texttt{RobinCar2}. In the following sections, we provide a brief description of the statistical methods and the options available in the \texttt{RobinCar} and/or \texttt{RobinCar2} implementations.

First, we describe the data notation that we use throughout for continuous, discrete, and time-to-event  outcomes. Consider a trial with treatment groups $a = 1, ..., k$, and $n$ individuals. For individual $i$, let $Y_{a,i}$ be the potential outcome under treatment $a$, $ X_{i}$ denote baseline covariates to be used in the analysis, and $ Z_{i}$ denote baseline categorical variables used for covariate-adaptive randomization (CAR) with $L$ joint levels. Throughout, ``strata'' refers specifically to the strata variable(s) $ Z_i$ used in covariate-adaptive randomization. For example, $ Z_i$ could be a discretized version of a continuous $ X_i$. The strata $ Z_i$ may overlap entirely, somewhat, or not at all with $ X_i$. We emphasize that $ X_i$ and $ Z_i$ are baseline variables: they must be unaffected by treatment assignment.

Based on $ Z_i$, patients are assigned to a treatment group, indicated by $A_i \in \{1, ..., k\}$. Let $n_a = \sum_{i=1}^{n} I(A_i=a)$ be te total number of individuals assigned to treatment group $a$. The RobinCar Family allows for the following types of covariate-adaptive randomization: simple randomization (no covariate information used), biased coin randomization, stratified permuted block \citep{zelenRandomizationStratificationPatients1974}, and Pocock and Simon's minimization \citep{pocockSequentialTreatmentAssignment1975}.

\begin{table}
\small\centering
\caption{Statistical methods available in the RobinCar Family, and variations in \texttt{RobinCar} and \texttt{RobinCar2}}\label{tab:robincar-methods}
\begin{tabular}{p{4cm} | p{4cm} p{4cm}}
Description & \texttt{RobinCar} & \texttt{RobinCar2} \\
\hline \\
AIPW for average treatment effect based on a generalized linear working model &
 \texttt{robincar\_glm}, also allows for machine learning working models in the function \texttt{robincar\_SL}, also considers \citet{pocockSequentialTreatmentAssignment1975}'s minimization &
 \texttt{robin\_glm}, includes influence function-based or sandwich-based (under simple randomization only) variance options \\
& & \\
\hline
& & \\
Calibrated AIPW estimators for guaranteed efficiency gain and universal applicability across CAR schemes & 
\texttt{robincar\_calibrate} which includes linear and joint calibration & N/A \\
& & \\
\hline
& & \\
Mantel-Haenszel (MH) estimates of difference in proportions & \texttt{robincar\_mh}, with inference for traditional MH  estimand and ATE estimand & N/A \\
& & \\
\hline
& & \\
Covariate-adjusted stratified log-rank test & \texttt{robincar\_logrank}  & \texttt{robin\_surv} \\
& & \\
\hline
& & \\
Cox score test with model-robust variance & \texttt{robincar\_coxph} & N/A \\
& & \\
\hline
& & \\
Covariate-adjusted marginal hazard ratio estimation and inference & \texttt{robincar\_covhr} & \texttt{robin\_surv} \\
& & \\
\hline
& & \\
\end{tabular}
\end{table}

\subsection{Augmented Inverse Probability-Weighted Estimators}\label{sec:aipw}

The goal in this section is to estimate the treatment group response means:
\begin{align}\label{eq:theta}
    \theta_a = E(Y_a) \quad  \theta = (\theta_1, ..., \theta_k)^T.
\end{align}
We can then use the vector $ \theta$ to construct treatment contrasts (like the average treatment effect for continuous endpoints, or risk difference, risk ratio, or odds ratio for binary endpoints) using the delta method with a differentiable function $f( \theta)$. The typical un-adjusted estimator for $\theta$ is the sample mean within each treatment group, i.e., the estimator for $\theta_a$ is $\widehat{\theta}_{\rm SM, a} = \frac{1}{n_a} \sum_{i:A_i=a} Y_{a,i}$. As an example, the un-adjusted estimator for the difference in sample means between treatment groups 1 and 2 is $\widehat{\theta}_{\rm SM, 2} - \widehat{\theta}_{\rm SM, 1}$.

Several covariate-adjusted methods for estimating $ \theta$ are available in the RobinCar Family, which we walk through in this section.
The \textit{augmented inverse probability weighted (AIPW)} estimator can be considered a general estimator for the vector $\theta$ utilizing a ``working model'' for the conditional mean of the outcome $Y_a$ given a covariate vector $X$. The AIPW estimator for each element $\theta_a$ is:
\begin{align}\label{eq:aipw}
	\widehat{\theta}_{\rm AIPW, a} = \frac1n \sum_{i=1}^{n} \widehat{\mu}_a(X_i) + \frac{1}{n_a} \sum_{i:A_i=a} \left(Y_{a,i} - \widehat{\mu}_a(X_i) \right)
\end{align}
where $\widehat{\mu}_a$ is an estimate of the working model \citep{robinsEstimationRegressionCoefficients1994, zhangImprovingEfficiencyInferences2008}. For example, this working model could be a logistic regression for $Y_a$ using covariates $ X$. We emphasize that this is a \textit{model-assisted} approach. Estimation and inference (provided the correct variance estimators are used) does not presume that the working model is correct. If a generalized linear model with a canonical link function (e.g., linear model, logistic regression, or Poisson regression) is used, \eqref{eq:aipw} reduces to simply $\frac1n \sum_{i=1}^{n} \widehat{\mu}_a(X_i)$, the familiar G-computation, or Oaxaca-Blinder estimator \citep{freedmanRegressionAdjustmentsExperimental2008, guoGeneralizedOaxacaBlinderEstimator2023}, which is highlighted by the FDA as an acceptable method of covariate adjustment for marginal treatment effects \citep{u.s.foodanddrugadministrationAdjustingCovariatesRandomized2024}. If a generalized linear model with a non-canonical link is used (like negative binomial regression), then the second term in \eqref{eq:aipw} ensures that the average conditional mean estimate matches the sample mean of the outcome within each treatment group. Importantly, $\widehat{\theta}_{\rm AIPW,a}$ is consistent for $\theta_a$, under any of the considered covariate-adaptive randomization schemes, and even if $\mu_a(X)$ is a poor representation of $E(Y_a|X)$ \citep{bannickGeneralFormCovariate2025}.

\texttt{RobinCar} and \texttt{RobinCar2} users can implement \eqref{eq:aipw} using a \texttt{glm} or a \texttt{MASS::glm.nb} (negative binomial) working model for $\mu_a(X)$. This working model can be arbitrary: it can include variables used in covariate-adaptive randomization, it can be treatment-group specific or be estimated using pooled treatment groups, and it can include any types of model specifications that are allowed with the \texttt{formula} of \texttt{glm}.

In some settings, there may be a high-dimensional covariate vector $X$, and/or limited knowledge about the best functional form to specify for how the covariates are related to the outcome $E(Y_a|X)$. An alternative option is to estimate $E(Y_a|X)$ using machine learning methods. \cite{bannickGeneralFormCovariate2025} provides technical details justifying the use of these highly-flexible models under CAR. \texttt{RobinCar} has a function called \texttt{robincar\_SL} which uses the SuperLearner package that can incorporate single machine learning models, or an ensemble of models in the estimation of $E(Y_a|X)$ (e.g., random forest, XGBoost, or LASSO regression). An important note when using these models is that they require \textit{cross-fitting}, which separates the dataset into multiple folds so that the observations used to estimate a particular $\widehat{\mu}_a(X_i)$ did not contain observation $i$ (as a result, the final form of the estimator is an average of fold-specific AIPW estimators) \citep{chernozhukovDoubleDebiasedNeyman2017}. This prevents against over-fitting, and can be customized with the option \texttt{k\_split} which specifies the number of folds.

\subsubsection{AIPW with ``Super-Covariates''}

The FDA and the European Medicines Agency (EMA) have commented on an approach to covariate adjustment from Unlearn.AI called Prognostic Covariate Adjustment (PROCOVA\texttrademark) \citep{EMA_PROCOVA_2022}. In this approach, one trains a model on historical data, and uses that model to generate a prognostic score for all trial participants based on their baseline covariates \citep{schuler_increasing_2022}. Their prognostic score can then be used as a single covariate -- referred to as a ``super covariate'' -- in ANCOVA. The RobinCar Family can accommodate this approach as long as the values for the super covariate are generated ahead of time. The idea of using a super covariate based on a large historical dataset is applicable not only in ANCOVA but also in the more flexible AIPW framework as well. In fact, all of the considerations in Section \ref{sec:aipw} apply whether one is using baseline covariates, or a super covariate trained on historical data and generated from baseline covariates.

\subsubsection{Considerations for Variance Estimation}

Particular care in using \eqref{eq:aipw} must be taken with regards to variance estimation. Our estimand is \eqref{eq:theta}, and the variance estimator must be appropriate for this marginal estimand. That is, it should reflect appropriate asymptotic performance unconditional on the covariates $X$ that are used. This is in contrast to a conditional estimand where inference with respect to the specific covariate profile observed in the trial at hand is desired \citep{magirr_estimating_2025}. Furthermore, the variance estimator needs to be robust when $\mu_a(X) \neq E(Y_a|X)$ and possibly account for covariate-adaptive randomization. R users interested in conditional estimands should consider using the package \texttt{beeca} \citep{przybylskiBeecaBinaryEndpoint2024}.

\paragraph{Variance Estimators}

\cite{bannickGeneralFormCovariate2025} provides a variance estimator for $\widehat{\theta}_{\rm AIPW}$ that meets these criteria, which is derived based on its influence function. Based on the asymptotic form of the variance, there are actually several estimators for the variance that one could consider. Under simple randomization (SR), the asymptotic variance is:
\begin{align}\label{eq:variance-sr}
	V_{\rm SR} &= \mathrm{diag}\left[\pi_a \mathrm{Var} \left\{Y_a - \mu_a(X) \right\}, a = 1, \dots, K \right] \nonumber \\
	&\quad\quad + \mathrm{Cov} \left\{Y - \mu(X), \mu(X) \right\} + \mathrm{Cov} \left\{\mu(X), Y\right\}
\end{align}
where $Y = (Y_1, \dots, Y_K)$ and $\mu(X) = (\mu_1(X), \dots, \mu_K(X))$. The standard variance estimator presented by \cite{bannickGeneralFormCovariate2025} estimates the diagonal terms by decomposing $\mathrm{Var}\{Y_a - \mu_a(X)\}$ into the variance and covariances of $Y_a$ and $\mu_a(X)$, and estimating any term with $\mathrm{Var}(\mu_a(X))$ or $\mathrm{Cov}(\mu_a(X), \mu_b(X))$ by using the entire distribution of $X$ across the trial (rather than only in groups $a$ and $b$). In \texttt{RobinCar}, this is the standard option (\texttt{variance\_type = 1}) because it provides good performance when there may be many zeroes in $Y_a$. Another option is to estimate $\mathrm{Var}\{Y_a - \mu_a(X)\}$ directly using the sample variance among individuals in group $a$ (\texttt{variance\_type = 2}). A final option is to use only $X$ from groups $a$ and $b$ when estimating the $\mathrm{Cov}(\mu_a(X), \mu_b(X))$ terms (\texttt{variance\_type = 3}), which is appealing if in the sample at hand, the distribution of $X$ varies considerably between treatment groups. Otherwise, there could be a large mismatch between the $\mathrm{Cov}(Y_a, \mu_a(X))$ terms, which can only ever be estimated with data from group $a$, and the $\mathrm{Var}(\mu_a(X))$ terms, which can be estimated using everyone in the trial. When sample sizes are not too small, the three variance estimators are consistent with one another. However, in extreme scenarios (e.g., rare binary endpoints), differences may arise in finite samples.

\texttt{RobinCar2} provides these first two options for variance estimation as well. Setting \\\texttt{vcov\_args=list(decompose=FALSE)} is equivalent to \texttt{variance\_type = 2}, otherwise the default is \texttt{variance\_type = 1}. Additionally, \texttt{RobinCar2} also has an option to use the sandwich variance estimator, also known as the Eicker-Huber-White (E-H-W) standard errors. This estimator is not the same as that in \cite{bannickGeneralFormCovariate2025}, and it is only valid when the $\mu_a(X)$ is a linear model without treatment-by-covariate interactions (i.e., $\mu_a(X) = \alpha_a + \beta^T X$ where the coefficient vector $\beta$ is shared between treatment groups), and simple randomization was used. Its advantage is that it has various finite-sample adjustments (HC1, HC2, and HC3) that may improve performance when sample sizes are small.

\paragraph{Efficiency Gain}

One important property of a covariate-adjusted estimator is \textit{efficiency gain}. That is, the asymptotic variance of the estimator is smaller than the asymptotic variance of the sample mean $\bar{Y}_a$. If the working model is correctly specified using $X$ (i.e., $\mu_a( X) = E(Y_a |  X)$), then the AIPW estimator achieves the semiparametric efficiency bound in simple randomization \citep{tsiatisCovariateAdjustmentTwosample2008}. However, if the working model is not correctly specified, the AIPW estimator remains consistent but may result in efficiency loss. \cite{cohenNoharmCalibrationGeneralized2024}  proposed an intuitive method called \textit{linear calibration} that can restore efficiency gain in this setting for simple randomization. Briefly, the estimated potential outcome vector $\widehat{\mu}(X_i) = (\widehat{\mu}_1(X_i), \dots, \widehat{\mu}_K(X_i))$ is used as the new covariate vector in a linear regression for $Y_{a,i}$ and the linear regression model with $\widehat{\mu}(X_i)$ is used as the final fitted model. This estimator is available in \texttt{RobinCar} using the function \texttt{robincar\_calibrate} with the option \texttt{joint=FALSE}.

\paragraph{Covariate-Adaptive Randomization}

Under covariate-adaptive randomization, the asymptotic variance is more complicated, making efficiency gain harder to achieve. Furthermore, in Pocock and Simon's minimization, it is possible that the variance is not even estimable because the form of the variance depends on unknown properties of the randomization scheme. Thus, a second property of a covariate-adjusted estimator is that it has \textit{universal applicability}. This means that its variance does not depend on the covariate-adaptive randomization scheme used, i.e., the variance estimator that is obtained under simple randomization based on the form in \eqref{eq:variance-sr} is valid under any randomization scheme.

For example, when using a generalized linear model with a canonical link function, the variance of \eqref{eq:aipw} is only estimable under Pocock and Simon's minimization if $Z$ are included in the covariate vector $X$ and treatment-by-covariate interactions in $\mu_a(X)$ are used. To address these challenges, \cite{bannickGeneralFormCovariate2025} propose an estimator called \textit{joint calibration}, which builds off of linear calibration by adding the additional $Z_i$ to the new covariate vector in the second-stage linear regression. The joint calibration estimator is given by:
\begin{align*}
    \widehat{\theta}_{\rm JC, a} &= \frac1n \sum_{i=1}^{n} \left\{\widehat{\alpha}_{a}^T Z_i +   \widehat{\beta}_a^T  \widehat {\mu} (X_i) \right\}\\
    (\widehat{\alpha}_{a}, \widehat {\beta}_a) &= \arg\min_{ \alpha_{a}, \beta_a  } \sum_{i:A_i=a} \left\{ Y_{a,i}  -  \alpha_{a}^T Z_i  -   \beta_a^T  \widehat {\mu} (X_i) \right\}^2
\end{align*}
where $Z_i$ is coded as a dummy vector specifying the stratification group for observation $i$.
This estimator for $\theta_a$ is always asymptotically more precise than the sample mean, and its variance does not depend on the randomization scheme used, bypassing the issues with Pocock and Simon's minimization. It is available using \texttt{robincar\_calibrate} with the option \texttt{joint=TRUE}. Table \ref{tab:universality-geg} summarizes these issues and the properties of the available methods in \texttt{RobinCar}.
 
 \begin{table}[]
 \small
    
    \caption{Universal applicability (U) and efficiency gain (EG) using a generalized linear model as the working model separately for each treatment group $a$. The columns indicate correct or incorrect specification of the conditional mean model, and whether strata are included as covariates in the working model. Special cases are denoted as follows. $\dagger$: using canonical link function; $\ddagger$: linear link function. Efficiency gain under simple randomization only is denoted as EG (SR). LC: linear calibration; JC: joint calibration.}
    \renewcommand{\arraystretch}{1.5}
    \begin{tabular}{c|c|c|c|c}
    & \multicolumn{2}{c|}{$Z \subseteq X$} & \multicolumn{2}{c}{$ Z \not\subseteq X$} \\
    \hline
    & $\mu_a(X) = E(Y_a |X)$ & $\mu_a(X) \neq E(Y_a | X)$ & $\mu_a( X) = E(Y_a |X)$ & $\mu_a(X) \neq E(Y_a | X)$ \\
    \hline
    \hline
        AIPW & U \& EG & U$^{\dagger}$ \& EG$^{\ddagger}$ & EG (SR) & -- \\
        \hline
        LC & U \& EG & \multicolumn{3}{c}{EG (SR)} \\
        \hline
        JC & \multicolumn{4}{c}{U \& EG} \\
        \hline
    \end{tabular}
    \label{tab:universality-geg}
\end{table}
 
\subsection{Robust Mantel–Haenszel Analysis for Binary Outcomes}

The Mantel-Haenszel (MH) risk difference estimator is widely used to perform for stratified analysis of binary endpoints. It has traditionally been viewed as estimating a common risk difference across strata. Recent work from \cite{qiuClarifyingRoleMantelHaenszel2025} showed the interesting result that under reasonable assumptions, the MH risk difference estimator also targets the Average Treatment Effect (ATE) estimand, defined as $$\delta_{\rm ATE} = E(Y_a) - E(Y_b) \equiv \theta_a - \theta_b.$$

In \texttt{RobinCar}, the MH estimator is applicable for a comparison of two treatment groups, when covariate adjustment using only categorical variables is desired, and under simple randomization. In this section, assume that the $X_i$ is a participants' categorical variable for covariate adjustment (which could be the joint levels of multiple categorical variables), which have a finite number of levels $L < \infty$. Let $n_a(l) = \sum_{i=1}^{n} I(X_i=l) I(A_i=a)$ be the number of participants assigned to treatment $a$ who have covariate category $l$. The MH estimator, first introduced in \cite{greenlandEstimationCommonEffect1985}, is a weighted-average of the category-specific effect estimates. The effect within a category $X_i = l$ is given by $\widehat{\delta}_l = \sum_{X_i=l} I(A_i=a) Y_i / n_a(l) - \sum_{X_i=l} I(A_i=b) Y_i / n_b(l)$. Their weighted average is given by:
\begin{align*}
	\widehat{\delta} = \frac{\sum_{l=1}^{L} w_l \widehat{\delta}_l}{\sum_{l=1}^{L} w_l} \quad w_l = \frac{n_a(l) n_b(l)}{n_a(l) + n_b(l)}.
\end{align*}

\cite{qiuClarifyingRoleMantelHaenszel2025} show that $\widehat{\delta}$ is a consistent and asymptotically normal estimator for $\delta_{ATE}$. Its main advantage is its applicability under both large-stratum and sparse-stratum regimes. \texttt{RobinCar} implements $\widehat{\delta}$ in the function \texttt{robincar\_mh}, also providing an appropriate variance estimator for the ATE estimand that is not available in other software. Even when the number of levels of the categorical variable $X_i$ is not fixed (the number of categories can grow with the sample size), $\widehat{\delta}$ is still a valid estimator for $\delta_{ATE}$ as long as there isn't extreme effect modification (see \cite{qiuClarifyingRoleMantelHaenszel2025} for technical assumptions). If there is, then $\widehat{\delta}$ is a consistent and asymptotically normal estimator for the estimand that is typically assumed when conducting a MH analysis: a weighted average of the category-specific treatment effects given by $\delta_{MH} = \sum_{l=1}^{L} w_l \delta_l / \sum_{l=1}^{L} w_l$, where $\delta_l = E(Y_a | X=l) - E(Y_b | X=l)$. However, this estimand is less interpretable and depends on the sizes of the strata. \texttt{RobinCar} still includes this estimand $\delta_{MH}$ (which impacts the variance formula) as an option for comparison in \texttt{robincar\_mh}.

\subsection{Survival Analysis for Time-to-Event Outcomes}

In this section, we explain the methods for covariate-adjusted (stratified) logrank tests and covariate-adjusted hazard ratios from \citet{yeCovariateAdjustedLogRankTest2023}, and the robust Cox score tests from \citet{yeRobustTestsTreatment2020}. Like the previous section, we focus on only two treatment groups at a time. Let $T_{i}$ and $C_{i}$ be the failure times and censoring times. Let $\Delta_{i} := T_{i} \leq C_{i}$ be an indicator of observing the event.
In this section, we are interested in conducting hypothesis tests of the difference in hazard between two treatment groups $a$ and $b$, and also to estimate marginal hazard ratios between the two groups.
\subsubsection{Hypothesis Testing}
 Let $D_a(t) = \sum_{i:A_i=a} I(T_i = t) \Delta_i$ be the total number of events observed at time $t$ in group $a$, and $R_a(t) = \sum_{i:A_i=a} I(T_i \geq t)$ be the number at risk at time $t$ in group $a$. Let $t_1 < \dots < t_K$ be a series of distinct event times observed across both treatment groups. The typical logrank test statistic is $T_L = \sqrt{n} \widehat{U}_L / \widehat{\sigma}_L$, where
\begin{align}\label{logrank}
    \widehat{U}_{L} = \sum_{k=1}^{K} \left[D_1(t_k) - R_1(t_k) \cdot \frac{D_1(t_k) + D_0(t_k)}{R_1(t_k) + R_0(t_k)}\right]
\end{align}
and $\widehat{\sigma}_L$ is the estimated standard deviation of the numerator. $T_L$ is compared to a standard normal distribution to test the hypothesis:
\begin{align}\label{eq:h0}
    H_0: \lambda_1(t) = \lambda_0(t) \quad \forall t.
\end{align}
where $\lambda_1(t)$ is the hazard function in treatment $A = 1$ and $\lambda_0(t)$ is the hazard function in treatment $A = 0$. \cite{yeCovariateAdjustedLogRankTest2023} shows that $\widehat{U}_L$ has a convenient asymptotically linear form which facilitates covariate adjustment for covariates $X$ using a linear model. Using this strategy, a new numerator $\widehat{U}_{CL}$ can be formed (which is the difference of the original $\widehat{U}_{L}$ and a mean-zero variance-reduction term), and it can be paired with an appropriate $\widehat{\sigma}_{CL} \leq \widehat{\sigma}_{L}$ for the denominator. This results in a covariate-adjusted logrank test statistic which does not change the null hypothesis \eqref{eq:h0}, and does not require any assumptions on censoring beyond what is already required for an un-adjusted logrank test (see Table \ref{tab:my_label}).

We use the notation $\lambda_a(t|X)$ to indicate the hazard function for group $a$ conditional on variables $X$. This is in contrast to $\lambda_a(t)$, which is the unconditional hazard function for group $a$.
The stratified log-rank test is a variation of the log-rank test that tests a stronger null hypothesis: that the hazard is the same between two treatment groups, and within some specific strata $Z$, i.e.,
\begin{align}\label{eq:H0strat}
	H_0: \lambda_1(t|z) = \lambda_0(t|z) \quad \forall(t, z).
\end{align}
 The stratified log-rank test is appropriate when covariate-adaptive randomization using variables $Z$ has been used in a clinical trial. \cite{yeCovariateAdjustedLogRankTest2023} provides a similar stratified covariate-adjustment strategy for \eqref{eq:H0strat} that can adjust for additional covariates (which is particularly useful for continuous covariates). \texttt{RobinCar} and \texttt{RobinCar2} implement these potentially stratified, covariate-adjusted logrank tests.
 
 Hypothesis testing for a difference in hazards can also be accomplished by using the score function of the Cox proportional hazards model, given by
\begin{align}\label{eq:cox}
    \lambda_1(t | X) &= \lambda_0(t|X) \exp(\theta) \\
    \lambda_0(t|X) &= h(t) \exp(\beta^T X)
\end{align}
where $h(t)$ is some unspecified baseline hazard function.
The null hypothesis in this case is $\theta = 0$, or equivalently:
\begin{align}\label{eq:h0cox}
    H_0: \lambda_0(t|X) = \lambda_0(t|X) \quad \forall t.
\end{align}
The null hypothesis explicitly depends on the covariates chosen for inclusion in the model $X$. A limitation of existing test statistics for the hypothesis in \eqref{eq:h0cox} is that they rely on correct specification of the Cox model. Specifically, the variance of the score that is used to standardize the test statistic is only valid under correct model specification. \cite{yeRobustTestsTreatment2020} show how to modify the denominator of the test statistic so that it does not rely on \ref{eq:cox} to be correct; it is instead treated as a working model. \texttt{RobinCar} implements this method so that practitioners interested in testing the null hypothesis in \eqref{eq:h0cox} have a robust alternative to the traditional Cox score test.


\begin{table}[]
    \small
    \centering
        \caption{Null hypotheses $H_0$ and censoring assumptions for the Robust Cox Score test, the logrank (L) and covariate-adjusted logrank (CL) tests, and the stratified logrank (SL) and covariate-adjusted stratified logrank (CSL) tests.}
    \label{tab:my_label}
    \begin{tabular}{c|c|c|c}
         & \multirow{2}{*}{$H_0$} & Censoring at & Valid Censoring \\
         & & Random Assumption &  Assumption \\
         \hline
         \hline
         & & & \\
        Robust & \multirow{2}{*}{$\lambda_1(t|X) = \lambda_0(t|X) $} &
        \multirow{2}{*}{$T_i \perp C_i | (A_i, X_i) $} & 
        $P(C_{i} \geq t | A_i=1, X_i) / P(C_{i} \geq t | A_i=0, X_i) $ \\
        Cox Score & & & is a function of only $t$ \\
        & & & \\
        \hline
        & & & \\
        L/CL & $\lambda_1(t) = \lambda_0(t) $ & $T_i \perp C_i | A_i$ & -- \\
        & & & \\
        \hline
        & & & \\
        SL/CSL & $\lambda_1(t|Z) = \lambda_0(t|Z)$ & $T_i \perp C_i | (A_i, Z_i)$ & -- \\
        & & & \\
        \hline
    \end{tabular}
\end{table}

\begin{table}[]
    \small
    \centering
        \caption{Comparison of the modeling assumption for the relationship between hazards in group $a = 1$ and $a = 0$, and their estimands, for the Cox proportional hazards model versus the covariate-adjusted estimator in \citet{yeCovariateAdjustedLogRankTest2023}.}
    \begin{tabular}{c|cc}
        & Model Assumption & Estimand \\
        \hline
        \hline
        & & \\
        Cox Model & $\lambda_a(t|X) = h(t)\exp(\theta \cdot a + \beta^T X)$ & $\exp(\theta) \equiv \lambda_1(t|X) / \lambda_0(t|X)$ \\
        & & \\
        \hline
        & & \\
        Estimator in &  
            \multirow{2}{*}{$\lambda_a(t) = h(t) \exp(\theta a)$} & 
            \multirow{2}{*}{$\exp(\theta) \equiv \lambda_1(t) / \lambda_0(t)$} \\
        \citet{yeCovariateAdjustedLogRankTest2023} & & \\
        & & \\
        \hline
    \end{tabular}
    \label{tab:surv-estimation}
\end{table}

\subsubsection{Estimating Marginal Hazard Ratios}

The $\theta$ from the Cox proportional hazards model in \eqref{eq:cox} is a hazard ratio conditional on the covariates included in the model. Although this is common-place, it is especially difficult to interpret when conditional on covariates because of non-collapsibility \citep{u.s.foodanddrugadministrationAdjustingCovariatesRandomized2024}. As a result, practitioners have desired a covariate-adjusted, \textit{marginal} hazard ratio. We contrast these two estimands in Table \ref{tab:surv-estimation}.

For the Cox score test, the method of estimation of the treatment effect is exactly what one would do using a Cox model. However, the interpretability of the estimated coefficients requires stronger assumptions than hypothesis testing: specifically, we need the Cox model in \eqref{eq:cox} to hold (this violates the RobinCar Family principles). \cite{yeCovariateAdjustedLogRankTest2023} show that we can instead perform estimation using the framework of the covariate-adjusted logrank test. In this case, we only need proportional hazards to hold marginally $\lambda_1(t) = \lambda_0(t) \exp(\theta)$, and do not rely on the exact form of \eqref{eq:cox} to be correct. This is what would be required for an un-adjusted analysis, i.e., using the Cox model with no covariates. In other words, we want to estimate $\theta$ using covariates to reduce the variance of $\widehat{\theta}$, but without having to correctly specify a model as in \eqref{eq:cox}. In essence, $\widehat{U}_{CL}$ can be treated as a variance-reduced score function for the Cox proportional hazards model with no covariates, and a covariate-adjusted $\theta$ can be obtained as the solution to this score function. There is an equivalent stratified version mirroring the stratified covariate-adjusted log-rank test \citep{yeCovariateAdjustedLogRankTest2023}. In this case, the estimand is $\exp(\theta) = \lambda_1(t|Z) / \lambda_0(t|Z) $.

\section{Case Study}

In this section, we provide examples of how to use the \texttt{RobinCar} and \texttt{RobinCar2} functions. We use data from the \cite{juraskaSpeff2trialSemiparametricEfficient2022} R package, which comes from a study of nucleoside treatment regimens for individuals with HIV-1 \citep{hammerTrialComparingNucleoside1996} called the AIDS Clinical Trials Group Study 175. The trial used stratified permuted block randomization for its four treatment arms, using an individual's history of antiretroviral use as the stratification variable (0 weeks of use, between 1-52 weeks of use, and 52+ weeks of use), named \texttt{strat}.

There are several outcomes of interest in the dataset, and possible estimands that could align with the trial objectives for those outcomes. Some of these outcomes are listed in Table \ref{tab:outcomes-example}. A full vignette showcasing the methods for all three of these outcomes is available at: \url{https://marlenabannick.com/RobinCar/articles/Family.html}. We will only discuss a subset of the results in the vignette in this section. All of the \texttt{RobinCar} functions can be used with multiple treatment groups. However, in this section, we subset to two treatment groups: zidovudine (0) and zidovudine plus didanosine (1). There are many covariates of interest in the dataset, but in our examples, we will focus on just three of them: having had previous non-zidovudine antiretroviral therapy (\texttt{oprior}), having hemophilia (\texttt{hemo}), and weight at baseline (\texttt{wtkg}). 

\begin{figure}[h]
\caption{Example of \texttt{RobinCar2} functions for covariate adjustment for binary outcomes (left) and time-to-event outcomes (right).}\label{fig:examples}
\vspace{0.1in}
\includegraphics[width=\textwidth]{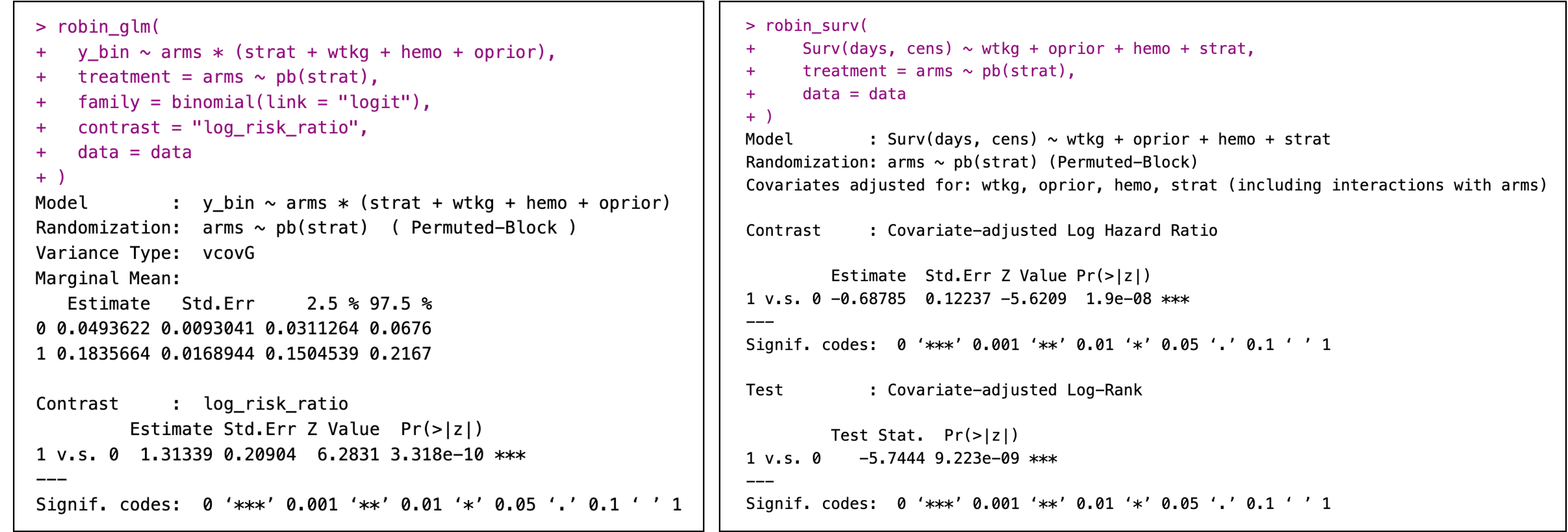}
\end{figure}

\texttt{RobinCar} and \texttt{RobinCar2} both have functionality to do covariate adjustment using the AIPW estimators described in Section \ref{sec:aipw}. Their implementation of the AIPW estimators closely mirrors the user interface of the \texttt{lm} and \texttt{glm} function in R. For exmaple, users familiar with the formula and family arguments of these functions will find \lm and \glm functions in \texttt{RobinCar2} to feel very natural. The left panel of Figure \ref{fig:examples} shows an example of using \texttt{RobinCar2} to obtain covariate-adjusted estimates of the risk reduction of CD4 decline $\geq 50\%$ between baseline and week 20 of follow-up, when didanosine is added to zidovudine. The working model has treatment-by-covariate interactions, and includes the stratification factor \texttt{strat} as a covariate. The treatment argument encodes how the treatment variable ``arms'' was generated: \texttt{pb(strat)} indicates stratified permuted block randomization was used based on the variable called \texttt{strat}. The output of the function call shows the estimates of the marginal means (this is the $\theta$ vector), and the desired contrast (risk ratio), along with standard errors, p-values, and confidence intervals.

Similarly, the \texttt{RobinCar} and \texttt{RobinCar2} packages both have functionality to do stratified and un-stratified, covariate-adjusted log-rank tests, and estimate marginal, covariate-adjusted hazard ratios based on the test chosen. The \texttt{RobinCar2} implementation closely mirrors the user interface of \texttt{survfit} or \texttt{coxph}, where the formula specification requires a \texttt{Surv(time, event)} on the left-hand side. The treatment argument is encoded exactly the same way as in the AIPW methods. The right panel of Figure \ref{fig:examples} shows an example of using \texttt{RobinCar2} to obtain covariate-adjusted estimates of the hazard ratio of the composite event of a CD4 decline of at least 50, an AIDS-defining event, or death, comparing didanosine + zidovudine to zidovudine alone. It also includes the test statistic and p-value for the covariate-adjusted logrank test. Since stratified permuted block randomization was used in this trial, the stratification factor must either be used as a covariate, or included as a stratification factor for a stratified log-rank test \citep{yeCovariateAdjustedLogRankTest2023}.

\begin{table}[]
    \small
    \centering
        \caption{Outcomes in the \cite{hammerTrialComparingNucleoside1996} and commonly-used estimands paired with the relevant \texttt{RobinCar} and \texttt{RobinCar2} functions.}\label{tab:outcomes}
    \begin{tabular}{>{\raggedright\arraybackslash}m{3.5cm} >{\raggedright\arraybackslash}m{2cm}|>{\raggedright\arraybackslash}m{2cm} >{\raggedright\arraybackslash}m{5.5cm}}
	Outcome & Type & Estimand & Functions \\
	\hline
	\hline
	& & & \\
	change in CD4 from baseline to week 20 & 
	continuous & 
	difference in means & 
	\texttt{RobinCar2::robin\_lm()} \\
	& & & \\
	\hline
	& & & \\
	\multirow{8}{3.5cm}{decline of CD4 count by $\geq 50\%$ from baseline to week 20} & 
	\multirow{8}{2cm}{binary} & 
	\multirow{4}{2cm}{difference in proportions} & 
	\texttt{RobinCar2::robin\_glm(family="binomial",} \\
	& & & \hspace{0.1in}\texttt{contrast="difference")} \\
	& & & \\
	& & & \texttt{RobinCar::robincar\_mh(estimand="ATE")}\\
	& & & \\
	& & & \\
	& & 
	\multirow{2}{2cm}{relative risk} & 
	\texttt{RobinCar2::robin\_glm(family="binomial",} \\
	& & & \hspace{0.1in}\texttt{contrast="log\_risk\_ratio")} \\
	& & & \\
    & & 
	\multirow{2}{2cm}{odds ratio} & 
	\texttt{RobinCar2::robin\_glm(family="binomial",} \\
	& & & \hspace{0.1in}\texttt{contrast="log\_odds\_ratio")} \\
	& & & \\
	\hline
	& & & \\
	\multirow{2}{3.5cm}{time until CD4 decline of 50, AIDS-defining event, or death} &
	\multirow{2}{2cm}{time-to-event} &
	\multirow{2}{2cm}{marginal hazard ratio} & 
	\texttt{RobinCar2::robin\_surv(test="logrank",} \\
	& & & \hspace{0.1in}\texttt{contrast="hazardratio")} \\
	& & & \\
		& & & \\
	\hline
	\end{tabular}
    \label{tab:outcomes-example}
\end{table}

\section{Conclusion}

The RobinCar Family provides a comprehensive suite of robust statistical methods for covariate adjustment in randomized clinical trials, in line with recent FDA guidance \citep{u.s.foodanddrugadministrationAdjustingCovariatesRandomized2024}. The packages \texttt{RobinCar} and \texttt{RobinCar2} fill an important gap, facilitating the practical implementation of powerful new statistical methods. Finally, \texttt{RobinCar2} is an open-source project of the American Statistical Association's Biopharmaceutical Section Covariate Adjustment Scientific Working Group. As such, it is supported by a community of developers across industry and academia, ensuring its longterm maintenance and sustainability.

\newpage
\section{Availability and Requirements}

\texttt{RobinCar} is available on CRAN with the following specifications:
\begin{itemize}
	\item \textbf{Project name:} RobinCar: Robust Inference for Covariate Adjustment in Randomized Clinical Trials
	\item \textbf{Project home page:} https://cran.r-project.org/web/packages/RobinCar/index.html
	\item \textbf{Operating system(s)}: Platform independent
	\item \textbf{Programming language:} R
	\item \textbf{Other requirements:} R version $\geq 2.10$
	\item \textbf{License:} MIT
	\item \textbf{Any restrictions to use by non-academics:} N/A
\end{itemize}

\noindent\texttt{RobinCar2} is available on CRAN with the following specifications:
\begin{itemize}
	\item \textbf{Project name:} RobinCar2: Robust Inference for Covariate Adjustment in Randomized Clinical Trials
	\item \textbf{Project home page:} https://cran.r-project.org/web/packages/RobinCar2/index.html
	\item \textbf{Operating system(s)}: Platform independent
	\item \textbf{Programming language:} R
	\item \textbf{Other requirements:} R version $\geq 3.6$
	\item \textbf{License:} Apache License 2.0
	\item \textbf{Any restrictions to use by non-academics:} N/A
\end{itemize}

\bibliographystyle{apalike}
\bibliography{references}

\end{document}